\documentclass[prl,twocolumn,superscriptaddress,showpacs,amsmath,amssymb,balancelastpage]{revtex4-1}

\pdfoutput=1

\makeatletter
\newcommand*{\balancecolsandclearpage}{%
	\close@column@grid
	\clearpage
	\twocolumngrid
}
\makeatother

\usepackage{dcolumn} % Align table columns on decimal point

\usepackage{xcolor,graphicx,mathtools,bm}

\definecolor{darkblue}{RGB}{0,0,150}
\definecolor{nightblue}{RGB}{0,0,100}

\usepackage[
colorlinks,
citecolor=darkblue,
linkcolor=darkblue,
urlcolor=nightblue]{hyperref}

\usepackage[english]{babel}
\usepackage[babel,kerning=true,spacing=true]{microtype}

\def\bk{{\bf k}}

\def\bq{{\bf q}}

\def\b0{{\bf 0}}

\def\Lam{\Lambda}

\begin{document}

\title{Anomalous dynamical scaling from nematic and U(1)-gauge field
 fluctuations in two dimensional metals}

\author{Tobias Holder}
\affiliation{Max-Planck-Institute for Solid State Research,
 D-70569 Stuttgart, Germany}
\author{Walter Metzner}
\affiliation{Max-Planck-Institute for Solid State Research,
 D-70569 Stuttgart, Germany}

\date{\today}

\begin{abstract}
We analyze the scaling theory of two-dimensional metallic electron systems in the presence of critical bosonic fluctuations with small wave vectors, which are either due to a $U(1)$ gauge field, or generated by an Ising nematic quantum critical point.
The one-loop dynamical exponent $z=3$ of these critical systems was shown previously to be robust up to three-loop order.
We show that the cancellations preventing anomalous contributions to $z$ at three-loop order have special reasons, such that anomalous dynamical scaling emerges at four-loop order.
\end{abstract}

\maketitle

A well-known mechanism for non-Fermi liquid behavior in two-dimensional
metals is provided by scattering at quantum critical bosonic degrees
of freedom.
The latter can be order parameter fluctuations at a quantum critical
point \cite{Loehneysen2007}, or emergent gauge fields arising from strong 
interactions \cite{Lee2006}.
The gauge field propagator diverges at small momenta and thus leads 
to singular forward scattering.
Order parameter fluctuations can be singular at small or large
momenta, depending on the nature of the phase transition.
Critical order parameter fluctuations with small momenta are present
at the onset of nematic order driven by a Pomeranchuk instability in
metallic electron systems \cite{Oganesyan2001,Metzner2003}.
In fact, the problem of two-dimensional fermions coupled to a
U(1)-gauge field emerging in doped Mott insulators is closely 
related to the problem posed by an Ising nematic quantum critical 
point (QCP) \cite{Metlitski2010}.

Both problems have a fairly long history. One-loop results for the
bosonic propagator and the fermion self-energy were first derived
in the context of the U(1)-gauge theory \cite{Lee1989}, and later
for the case of a nematic QCP \cite{Oganesyan2001,Metzner2003}.
The bosonic propagator is substantially modified by Landau damping, 
and the fermion self-energy scales as $|\omega|^{2/3}$ at low 
excitation energies, implying a pronounced non-Fermi liquid behavior
without Landau quasi-particles.
In case of an Ising nematic on a lattice, a momentum dependent 
form factor in the self-energy leads to a few ''cold spots'' on 
the Fermi surface, where quasi-particles survive \cite{Metzner2003}.
The main contributions to the self-energy at Fermi momenta $\bk_F$ 
come from particle-hole excitations near $\bk_F$ and $-\bk_F$ with 
a small momentum transfer $\bq$ that is almost tangential to the Fermi 
surface, and an excitation energy of the order $|\bq|^3$. 
At the one-loop level, both bosonic and fermionic degrees of freedom
obey scaling with a dynamical exponent $z=3$.

Over many years, the one-loop result was expected to be robust.
It was believed to be controlled by a an expansion in the inverse 
fermion flavor number $N_f$ \cite{Polchinski1994}, and at the two-loop
level no qualitative modifications were found \cite{Altshuler1994}.
Hence, it came as a surprise when Sung-Sik Lee \cite{Lee2009} discovered that the naive $1/N_f$-expansion is not valid, and Feynman diagrams of arbitrary loop order contribute even in the limit $N_f \to \infty$.
Shortly afterwards, Metlitski and Sachdev \cite{Metlitski2010}
formulated a general scaling theory for the nematic QCP and the
related U(1)-gauge field problem. Focusing on dominant
processes near $\bk_F$ and $-\bk_F$ they derived a low energy
field theory which involves only states near those two Fermi points.
Symmetry constraints allow for only two independent anomalous scaling
exponents, a fermion anomalous dimension $\eta_f$ and an anomalous
dynamical exponent $z \neq 3$.
A small contribution to $\eta_f$ was found by Metlitski and Sachdev in
a three-loop calculation of the fermion self-energy.
A renormalization of $z$ might be obtained from a divergence of
the boson self-energy, but that quantity was found to be finite
up to three-loop order.
These results were confirmed in subsequent extensions of the problem
which were designed such that the loop expansion corresponds to an
expansion in a suitably designed small parameter 
\cite{Mross2010,Dalidovich2013}.

\looseness=-1
Thus, the most important remaining issue is whether anomalous dynamical scaling appears at higher order or not, which has strong implications for observable quantities such as the compressibility \cite{Metlitski2010}.
In the following we clarify that question. It turns out that the absence of a renormalization of $z$ at three-loop order is due to cancellations
which are specific to that order, and cannot be expected to hold at
higher orders. We identify a divergent contribution to the boson self-energy at \emph{four-loop} order which cannot be cancelled by other contributions, and will thus lead to anomalous dynamical scaling with $z \neq 3$.

Our analysis is based on the effective field theory for the low energy behavior derived by Lee \cite{Lee2009,Lee2008} and Metlitski and Sachdev
\cite{Metlitski2010}. Focusing on the dominant excitation processes near a Fermi point $\bk_F$ and its antipode $-\bk_F$, and discarding irrelevant terms, one obtains an effective low-energy theory described by the Lagrangian \cite{Metlitski2010}
\begin{align}
	L &= \sum_{s=\pm} \psi^\dagger_s
	\left( \eta\partial_\tau -
	is\partial_x - \partial^2_y \right) \psi_s \notag\\
	& \quad -\sum_{s=\pm} g_s\phi\psi^\dagger_s\psi_s
	- \frac{N_f}{2e^2}(\partial_y\phi)^2 
	+ \frac{N_f}{2} r \phi^2 . 
\label{metlagrangian}
\end{align}
Here $\phi$ is a bosonic scalar field, while $\psi_{\pm}$, $\psi_{\pm}^{\dagger}$ are Grassmann fields with $N_f$ flavor components corresponding to fermionic excitations in the two "patches" near $\pm\bk_F$. 
In the U(1)-gauge field problem, $\phi$ is the transverse gauge field and $g_+ = -g_-$. For the Ising nematic, $\phi$ is the order parameter field and $g_+ = g_-$.
In both cases the physical flavor number is $N_f = 2$.
The derivatives are with respect to real space and imaginary time variables. The spatial coordinates have been chosen such that the corresponding momentum variables $k_x$ and $k_y$ are normal and tangential to the Fermi surface at $\pm\bk_F$, respectively.
Several numerical prefactors have been absorbed by a rescaling of fields and space coordinates. In particular $|g_s| = 1$.
The U(1)-gauge field is always massless so that $r=0$. For the Ising nematic case,
$r$ is generally finite, but vanishes at the QCP.

In random phase approximation (RPA), which corresponds to a one-loop calculation of the boson and fermion self-energies, the boson and fermion propagators at criticality ($r=0$) have the form \cite{Lee1989}
\begin{align}
	D^{-1}(q) &= 
	N_f \left( \frac{q_y^2}{e^2} + 
	\frac{1}{4\pi} \frac{|q_0|}{|q_y|} \right), \\
	G^{-1}_s(k) &=
	s k_x + k_y^2 - i \frac{\kappa}{N_f} \frac{k_0}{|k_0|^{1/3}} ,
\end{align}
with $\kappa = 2 e^{4/3}/[\sqrt{3}(4\pi)^{2/3}]$.
These propagators solve the RPA equations also self-consistently \cite{Polchinski1994}.
Note that the linear frequency term (proportional to $\eta$) in the fermion propagator is subleading compared to the self-energy and has therefore been discarded.
The RPA solution describes a non-Fermi liquid with fermions which are strongly scattered at overdamped bosons.
Both propagators are homogeneous under the scaling
\begin{align}
	q_y \to \lambda q_y , \qquad 
	q_x \to \lambda^2 q_x , \qquad
	q_0 \to \lambda^z q_0
\label{scaling}
\end{align}
with a dynamical exponent $z=3$.
Scaling the fields accordingly by a factor $\lambda^2$, the time derivative in the Lagrangian (\ref{metlagrangian}) is irrelevant, the boson mass term is relevant, while all other terms are marginal.

Metlitski and Sachdev \cite{Metlitski2010} have derived a general scaling ansatz for $D(q)$ and $G_s(k)$.
Ward identities following from symmetries of the low-energy theory constrain the renormalization of the marginal terms in the Lagrangian such that only two independent renormalizations are possible: a rescaling of the fermion field $\psi=Z_\psi^{1/2}\psi_r$ and a renormalization of the coupling constant $e^2 = Z_e e^2_r$. The former yields an anomalous fermionic scaling dimension, the latter an anomalous dynamical exponent.
In the framework of the field theoretical renormalization group, anomalous infrared scaling can be linked to ultraviolet (UV) divergences of the theory \cite{Amit1984}.
The anomalous scaling dimensions are thus given by \cite{Metlitski2010}
\begin{align}
	\eta_\psi &= 
	- \Lambda\frac{\partial}{\partial \Lambda}\log Z_\psi , \\
	\eta_e &= 3-z = 
	\Lambda\frac{\partial}{\partial \Lambda}\log Z_e ,
\end{align}
where $\Lambda$ is a UV cutoff restricting momenta and frequencies to $|q_y| \leq \Lambda$, $|q_x| \leq \Lambda^2$, and $|q_0| \leq \Lambda^3$.
The anomalous dimensions determine the scaling behavior of physical quantities.
For example, the fermion self-energy on the Fermi surface scales as 
$|\omega|^{(2-\eta_\psi)/z}$, and the fermionic density of states as 
$\omega^{\eta_\psi/z}$. In one-loop approximation one has $\eta_\psi = 0$ and $\eta_e = 0$, that is, $z=3$.

\begin{figure}[b]
\begin{center}
\includegraphics{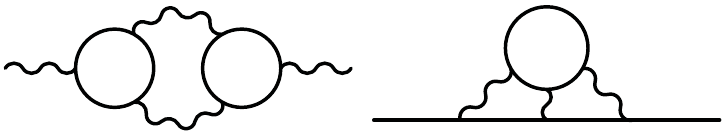}
\end{center}
\caption{Three-loop contributions to the boson (left) and fermion (right) self-energies computed by Metlitski and Sachdev \cite{Metlitski2010}.
The solid lines represent fermions, the wiggly lines bosons.}
\label{fig:metnew}
\end{figure}

Calculations beyond the one-loop approximation are generally performed by expanding around the RPA solution, that is, by inserting RPA propagators for the internal lines in Feynman diagrams representing higher order contributions \cite{Altshuler1994,Lee2009,Metlitski2010}. 
This corresponds to a resummation of terms of arbitrary order. It reduces the infrared divergences and the number of diagrams contributing in a given loop order. However, the integrations are complicated by the non-rational frequency dependence of the fermionic RPA propagator.
Since the one-loop fermion self-energy depends only on frequency, not momentum, Ward identities are still valid order by order at least in the zero frequency limit \cite{Metlitski2010}.
Metlitski and Sachdev computed the anomalous dimensions from the fermion and boson self-energies at zero frequency up to three-loop order
\cite{Metlitski2010}. At two-loop order, no contributions were found. At three-loop order (see Fig.~\ref{fig:metnew}), a small contribution to the fermion anomalous dimension, $\eta_{\psi} \approx \pm 0.068$ for $N_f=2$, was discovered, with a plus (minus) sign for the nematic (gauge field) system. 
No contribution to $\eta_e$ was found up to three-loop order. Divergent contributions to the boson self-energy obtained from individual Feynman diagrams (of Aslamasov-Larkin type, see Fig.~\ref{fig:metnew}) cancel each other such that the sum is finite. It remained open whether similar cancellations occur at higher orders.

\begin{figure}[tb]
\begin{center}
\includegraphics{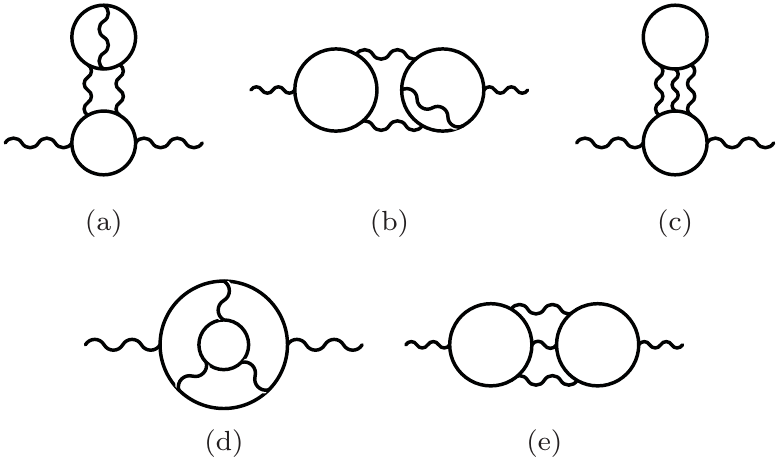}
\end{center}
\caption{Four-loop contributions to the boson self-energy. The diagrams do not represent all variants which can be produced by symmetrizing the fermion loops.}
\label{fig:fourclasses}
\end{figure}

We now discuss the general structure of contributions to the anomalous scaling dimensions. 
The fermionic anomalous dimension $\eta_\psi$ is obtained from the logarithmic UV divergence of 
$Z_\psi = 1 - \partial\Sigma_s(k)/\partial k_s$, where $\Sigma_s(k)$ is the fermion self-energy and $k_s = s k_x + k_y^2$.
The anomalous dynamical scaling dimension is determined by a logarithmic UV divergence of
$Z_e = 1 + \frac{e^2}{2N_f} \partial^2\Pi(q)/\partial q_y^2$, where $\Pi(q)$ is the boson self-energy.
The various contributions to the self-energies can be classified by the number of loop integrals.
Feynman diagrams representing four-loop contributions to the bosonic self-energy are shown in Fig.~\ref{fig:fourclasses}.

A major simplification concerns contributions where only fermions from one patch are involved. Analyzing the complex poles of the propagators,
Lee~\cite{Lee2009} proved that all planar one-patch corrections to the RPA self-energies are finite. For the boson self-energy, planar one-patch diagrams are even zero in the static limit \cite{Lee2009}. These observations are actually valid for one-patch contributions from non-planar diagrams, too.
Hence, one-patch corrections do not change the one-loop fixed point structure at any finite order.

The Feynman diagrams contain \emph{fermion loops} which are connected by boson lines. Feynman diagrams for the fermionic self-energy and the fermion-boson vertex contain also an open fermion line, but at least one fermion loop is required to have fermions on both patches mixed in one diagram.
We denote a fermion loop with $N$ vertices by $\Pi_{N,s}(q_1,\dots,q_N)$, where $s$ is the patch index and $q_1,\dots,q_N$ are the bosonic momenta injected at the vertices.
The sum over all permutations of vertices at a loop defines the \emph{symmetrized loop} $\Pi_{N,s}^{\rm sym}(q_1,\dots,q_N)$.
Symmetrized fermion loops can be viewed as effective interactions between bosons.
In the present problem, these effective interactions are singular (non-local) and marginal for any $N$ \cite{Thier2011}.
The Feynman diagrams can be grouped in distinct classes by writing them in terms of symmetrized loops.
Two diagrams within a class are related by a permutation of boson vertices attached to a fermion loop.
For example, the diagrams (c) and (d) in Fig.~\ref{fig:fourclasses} are related by such a permutation.
From diagram (e) one can generate several other diagrams within the same class by permuting the vertices attached to one of the loops.

Cancellations of divergences may occur between Feynman diagrams within a class defined by symmetrized loops. For bare fermion propagators substantial generic cancellations of infrared divergences upon symmetrization of a single fermion loop where found already long ago \cite{Neumayr1998,Kopper2001}.
However, the symmetrized loop remains singular in the very special infrared limit Eq.~(\ref{scaling}), where energies vanish much faster than momentum transfers and the latter become collinear \cite{Metlitski2010,Thier2011}.
The UV divergences found by Metlitski and Sachdev \cite{Metlitski2010} in Aslamasov-Larkin type contributions (see Fig.~\ref{fig:metnew}) to the bosonic self-energy cancel when adding the two distinct terms in the class of diagrams grouped by symmetrized loops.
On the other hand, there is no reason to expect systematic cancellations between diagrams not belonging to the same symmetrized loop class. Moreover, the relative sign of contributions from different classes depends on the signs of the fermion-boson vertices. For example, the product of couplings $g_s$ in the diagrams (c) and (d) in Fig.~\ref{fig:fourclasses} yields a minus sign in case of the U(1)-gauge theory, if the loops are on distinct patches, but a plus sign for the nematic QCP, while the product of couplings in diagrams of type (e) is always positive. Hence, even if a cancellation occured between distinct classes for one theory, it would not occur for the other.

The issue is thus under which circumstances and to what extent cancellations suppress symmetrized fermion loops in the ultraviolet limit.
Explicit expressions for $\Pi_{N,s}(q_1,\dots,q_N)$ constructed with RPA propagators are presented in the Supplementary Materials~\cite{citesuppl}. The energy-momentum integration within the fermion loop is convergent both in the infrared and ultraviolet limits. 
Naive power counting predicts that $\Pi_{N,s}(q_1,\dots,q_N)$ and $\Pi_{N,s}^{\rm sym}(q_1,\dots,q_N)$ scale as $\Lambda^{-2(N-3)}$ for $q_1,\dots,q_N$ tending to infinity as $q_{iy} \sim \Lambda$, $q_{ix} \sim \Lambda^2$, $q_{i0} \sim \Lambda^3$.
In particular, the 3-point loop is thus expected to become scale invariant for large $\Lam$.
However, the 3-point loop entering the Aslamasov-Larkin-type contribution to the boson self-energy actually behaves differently.
Since the external momentum $q$ of the boson self-energy stays fixed, the UV limit has to be taken only at the vertices of the loop connected to internal boson lines. A large momentum is injected at one of the vertices and pulled out again at a neighboring vertex.
From the explicit expression for the 3-point loop it is easy to see that $\Pi_{3,s}^{\rm sym}(q,p,-q-p)$ decays as $\Lambda^{-1}$ for large $p$ and fixed $q$. Moreover, $p_y$ is constrained to the size of $q_y$.
As a consequence, the symmetrized Aslamasov-Larkin diagram is finite in the UV limit \cite{Metlitski2010}.
The decay of the symmetrized 3-point loop with $\Lambda^{-1}$ is a consequence of the symmetrization, while the kinematic constraint of $p_y$ is present already in the unsymmetrized function $\Pi_{3,s}(q,p,-q-p)$.

\begin{figure}[htb]
\begin{center}
\includegraphics{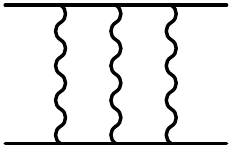}
\end{center}
\caption{The necessary building block for a diagram to be singular.}
\label{fig:buildblock}
\end{figure}

The gain of a power in $\Lambda^{-1}$ for a symmetrized loop with a fixed external momentum holds generally. In the Supplementary Materials~\cite{citesuppl} we show that $\Pi_{N,s}^{\rm sym}(q_1,\dots,q_N)$ vanishes if one of the momenta $q_i$ vanishes. This implies that the symmetrized loop decays at least as $\Lambda^{-2(N-3)-1}$ if one momentum stays fixed while the others tend to infinity.
If a fixed momentum $q$ is injected at a vertex and extracted at another vertex of the same loop, while all other external momenta become large (this is possible only for $N \geq 4$), there is even a suppression of order $\Lambda^{-2}$.
This gain in power-counting from symmetrization guarantees that contributions to the boson self-energy diverge at most logarithmically, with a prefactor of the order $q_y^2$ for $q_0 = 0$.
Quadratic and linear UV divergences of the boson self-energy indicated by power-counting must cancel. The same conclusion can be drawn from a Ward identity following from current conservation \cite{Metlitski2010}.
On the other hand, the kinematic constraint suppressing the 3-point loop in the Aslamasov-Larkin diagram is more special.
It does hold for $N$-point loops with general $N$, too, but only in case that only \emph{one} large bosonic momentum passes through the loop, that is, when a large momentum is injected into the loop at one of the vertices and extracted again at another, while the remaining $N-2$ momenta $q_i$ remain finite \cite{Holder2015}.
However, this restriction does not apply if a fermion loop is connected to a fermion line (open or closed) by three or more boson propagators, as in Fig.~\ref{fig:buildblock}.
For such diagrams, there is no suppression of the UV divergence from kinematic constraints.
For the fermionic self-energy, a fermion loop connected to another fermion line by at least three boson propagators appears already at three-loop order (see Fig.~\ref{fig:metnew}). This is the logarithmically divergent contribution to $Z_\psi$ identified by Metlitski and Sachdev \cite{Metlitski2010}.
For the boson self-energy, fermion lines connected by three boson lines as in Fig.~\ref{fig:buildblock} appear only at four-loop order.
Examples are the diagrams (c), (d) and (e) in Fig.~\ref{fig:fourclasses}.
Those diagrams can thus be expected to contribute to a logarithmic divergence of $Z_e$, and thus to an anomalous dynamical exponent.

The computation of four-loop diagrams is difficult.
To see that anomalous dynamical scaling indeed emerges at four-loop order, we have evaluated the sum of all diagrams in the symmetry class of the diagram (e) in Fig.~\ref{fig:fourclasses}. These are all diagrams where two four-point loops are connected by three boson propagators. There are six topologically distinct such diagrams.
The loop-integration is tricky due to the singular structure of the integrands. Details of the evaluation are presented in the Supplementary Materials~\cite{citesuppl}.
Summing all contributions, we expect a logarithmic UV divergence of the form
\begin{align}
 \Pi^{(4e)}(q) &\sim 
 C^{(4e)} \frac{q_y^2}{e^2} \log\left( \Lambda/|q_y| \right)
\label{log4e}
\end{align}
for $q_0 = q_x = 0$ and $\Lambda \gg |q_y|$.
Our calculation clearly yields a UV divergence with a negative prefactor.
A cancellation or kinematic constraint removing the divergence does not occur.
However, it turns out that $\Pi^{(4e)}(q)$ diverges actually as $[\log(\Lam_y/|q_y|)]^5$.
The reason for this unexpected stronger divergence is the $q_x$-independence of the boson propagator $D(q)$.
Adding tentatively a term proportional to $(q_x/q_y)^2$ to the denominator of $D(q)$, one recovers the expected simple log-divergence. For example, choosing the prefactor of that term as $1/e^2$, we obtain a logarithmic divergence with a prefactor $C^{(4e)} = -0.04$ for $N_f = 2$, yielding a small negative contribution $\eta^{(4e)} = -0.02$ to the anomalous scaling exponent.

A term of order $(q_x/q_y)^2$ has the same scaling dimension as the other terms in the boson propagator. On the other hand, it is not present in the RPA propagator, Eq.~(2), and it violates a ''reparametrization symmetry'' \cite{Metlitski2010} of the Lagrangian, Eq.~(1). However, that symmetry has so far been derived only for the classical Lagrangian and might be spoiled by a quantum anomaly. Alternatively, a $q_x$-dependence of the boson propagator might be generated by spontaneous symmetry breaking.
In any case, with a boson propagator depending only on $q_y$, the theory seems to be unrenormalizable.

\looseness=-1
The four-loop diagrams of type (e) do not contain any divergent subdiagrams from vertex or self-energy insertions.
The vertex corrections contained as subdiagrams in those diagrams include a 4-point loop with a fixed external boson momentum. The symmetrized vertex correction obtained by summing all permutations of boson vertices at the loop is thus finite due to the cancellations under symmetrization discussed above.
The divergence in the sum of diagrams of type (e) is thus a \emph{primitive} divergence which is not a consequence of the divergent fermionic renormalization $Z_{\psi}$.
By contrast, the diagram (c) in Fig.~\ref{fig:fourclasses} contains a divergent self-energy insertion, and diagram (d) a vertex correction which is most likely divergent, too. These divergences have to be compensated by counterterms, to disentangle them from UV divergences contributing to $Z_e$.

In summary, we have shown that the quantum field theory describing the Ising nematic QCP and non-relativistic electrons coupled to a $U(1)$-gauge field in two dimensions acquires anomalous dynamical scaling at four-loop order.
After the unexpected discovery of a fermionic anomalous dimension at three-loop order by Metlitski and Sachdev \cite{Metlitski2010}, this establishes another significant deviation from the at first sight robust one-loop result in that theory.
The four-loop contribution considered in our work tends to increase the dynamical scaling exponent $z$ to a value above three.
A surprising result, which requires further investigations, is that the boson self-energy computed with RPA propagators exhibits a non-renormalizable divergence.

%%%%%%%%%%%%%%%%%%%%%%%%%%%%%%%%%%%%%%%%%%%%%%%%%%%%%%%%%%%%%%%%%%%%%%%%

\begin{acknowledgments}

We are very grateful to D.~Belitz, A.~Chubukov, A.~Eberlein, H.~Gies, M.~Metlitski, S.~Sachdev, B.~Obert, and H.~Yamase for valuable discussions.

\end{acknowledgments}

%%%%%%%%%%%%%%%%%%%%%%%%%%%%%%%%%%%%%%%%%%%%%%%%%%%%%%%%%%%%%%%%%%%%%%

%

\balancecolsandclearpage

%%% Supplementary Material %%%%%%%%%%%%%%%%%%%%%%%%%%%%%%%%%
%%% Prefix "S" to all equations, figures, tables and reset the counter 

\setcounter{equation}{0}
\setcounter{figure}{0}
\setcounter{table}{0}
\setcounter{page}{1}
\makeatletter
\renewcommand{\theequation}{S\arabic{equation}}
\renewcommand{\thefigure}{S\arabic{figure}}
\renewcommand{\thetable}{S\arabic{table}}
\renewcommand{\bibnumfmt}[1]{[S#1]}
\renewcommand{\thesection}{\Alph{section}}

\begin{NoHyper}

\onecolumngrid

\section{Anomalous dynamical scaling from nematic and U(1)-gauge field
 fluctuations in two dimensional metals: Supplementary material}
\begin{center}
\vspace*{-6pt}T. Holder and W. Metzner
\end{center}

In this Supplementary Material we provide an explicit expression for the fermion loops. We prove that symmetrized loops vanish when one of the external energy-momentum variables vanishes, and we present details on the calculation of the four-loop contribution to the bosonic self-energy.

%%%%%%%%%%%%%%%%%%%%%%%%%%%%%%%%%%%%%%%%%%%%%%%%%%%%%%%%%%%%%%%%%%%%%%

\subsection{Explicit expression for N-point loops}

The $N$-point fermion loop on patch $s$ is defined by an integrated product of $N$ fermion propagators as
\begin{align}
 \Pi_{N,s}(q_1,\dots,q_N) = 
 I_{N,s}(p_1,\dots,p_N) =
 N_f \int \frac{dk_0}{2\pi} \int \frac{d^2k}{(2\pi)^2} 
 \prod_{j=1}^N G_s(k-p_j) \, .
\label{npointdef}
\end{align}
The energy-momentum variables $q_j$ and $p_j$ are related by
$q_j = p_{j+1} - p_j$ for $j=1,\dots,N-1$, and $q_N = p_1 - p_N$.
Note that $q_1 + \dots + q_N = 0$ due to energy and momentum conservation.
A Feynman graph representing a fermion loop is shown in Fig.~\ref{fig:npoint}.
\begin{figure}[htb]
\begin{center}
\includegraphics{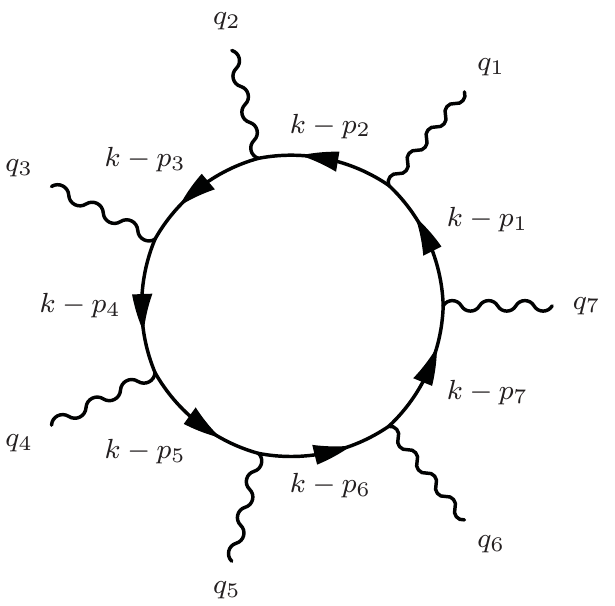}
\end{center}
\caption{$N$-point fermion loop for the case $N=7$. Solid lines represent fermion propagators, while the wiggly lines indicate the bosonic leg of the fermion-boson vertices.}
\label{fig:npoint}
\end{figure}

Since we expand around the one-loop (RPA) solution, the fermion propagators in the loop have the form
\begin{align}
 G_s^{-1}(k) = s k_x + k_y^2 -
 i \tilde\kappa \frac{k_0}{|k_0|^{1/3}}
\end{align}
with $\tilde\kappa = \kappa/N_f$.

The $k_x$ and $k_y$ integrations in Eq.~(\ref{npointdef}) can be easily done by residues. The $N$-point loop can then be written in the form
\begin{align}
 I_{N,s} = \frac{N_f}{2} \sum_{i<j}
 \int_{p_{i0}}^{p_{j0}} \frac{dk_0}{2\pi} \,
 \Theta\left( \frac{p_{i0}-p_{j0}}{p_{iy}-p_{jy}} \right)
 (p_{iy}-p_{jy})^{N-3}
 \prod_{k\neq i,j} J_{ijk,s}(k_0) \, ,
\label{loopform}
\end{align}
for $N \geq 3$, where
\begin{align}
 J_{ijk,s}(k_0) =
 \frac{1}{sD_{ijk} + F_{ijk}+i\Omega_{ijk}(k_0)} \, ,
\end{align}
with
\begin{align}
 D_{ijk} &= p_{ix}(p_{ky}-p_{jy}) + {\rm cycl}, \\
 F_{ijk} &= (p_{jy}-p_{iy})(p_{ky}-p_{jy})(p_{iy}-p_{ky}), \\
 \Omega_{ijk}(k_0) &=
 \tilde\kappa\frac{k_0-p_{i0}}{|k_0-p_{i0}|^{1/3}}(p_{ky}-p_{jy}) 
 + {\rm cycl} \, .
\end{align}
Here "cycl" denotes cyclic permutations of the indices $i,j,k$.

Note that the $k$-integration converges both in the infrared and ultraviolet limits, such that the $N$-point loop is a cutoff-independent function of the external energy-momentum variables.
Under a rescaling of the form $q_{iy} \to \lambda q_{iy}$, 
$q_{ix} \to \lambda^2 q_{ix}$, $q_{i0} \to \lambda^3 q_{i0}$, the $N$-point loop scales homogeneously as $\lambda^{2(3-N)}$.

Constructing the loop with bare instead of RPA propagators, one obtains the same expression (\ref{loopform}), with $\Omega_{ijk}(k_0)$ replaced by the $k_0$-independent function
$\Omega_{ijk} = -p_{i0} (p_{ky} - p_{jy}) + {\rm cycl}$.
The $k_0$-integration can then be performed analytically.
The function $\Omega_{ijk}$ is subleading in the scaling limit compared to $D_{ijk}$ and $F_{ijk}$. 
Summing over the two patches the result for the $N$-point loop with bare propagators agrees with an earlier result derived by performing the scaling limit \emph{after} the $k$-integration \cite{SThier2011}.

\subsection{Reduction for symmetrized $N$-point loops}

Here we show that the symmetrized $N$-point loop vanishes, if one of the external momenta vanishes.
The symmetrized $N$-point loop is given by a sum over all permutations of external momenta.
For a vanishing external momentum two fermion lines in the loop carry the same internal momentum. By permutations the doubled fermion line is cycled around the loop.
Let us consider the symmetrized 4-point loop as an example.
Without loss of generality, we can chose $q_4$ as vanishing and sum the three permutations which are cyclic in $q_1,q_2,q_3$. The sum is given by
\begin{align}
 \int \frac{dk_0}{2\pi} \int \frac{d^2k}{(2\pi)^2} \,
 G_s(k-p_1)G_s(k-p_2)G_s(k-p_3)
 \left[G_s(k-p_1)+G_s(k-p_2)+G_s(k-p_3)\right],
\end{align}
where we used that $q_4=p_1-p_4=0$.
Since the denominator of $G_s(k-p_i)$ is linear in $k_x$, the integrand can be written as a $k_x$-derivative,
\begin{align}
 - s \frac{\partial}{\partial k_x}
 \left[G_s(k-p_1)G_s(k-p_2)G_s(k-p_3)\right].
\end{align}
Performing the $k_x$-integration one thus finds that the 4-point loop vanishes if one leg has a vanishing momentum.
The extension to arbitrary N is straightforward, the sum of all cyclic permutations of $N-1$ legs, while one leg with zero momentum is kept separate, will cancel exactly.
By dimensional analysis, the UV scaling of the symmetrized loop with one fixed external momentum is thus reduced by a factor $\Lambda^{-1}$.

Using analyticity and the invariance under $q_i \mapsto -q_i$ one can conclude that symmetrized loops vanish even quadratically, if a vanishing momentum $q$ enters and leaves the same loop at two distinct vertices, provided that the other momenta remain finite. Due to momentum conservation this is possible only for $N \geq 4$. Hence, the UV scaling of symmetrized loops for $N-2$ large momenta and two fixed external momenta $q$ and $-q$ is reduced by a factor $\Lambda^{-2}$.

\subsection{Calculation of the four-loop analogue of the Aslamazov-Larkin diagram}

The four-loop contributions of type (e), where two 4-point fermion loops are connected by three boson propagators, are given by
\begin{align}
 \Pi^{(4e)}(q) & = -2 g_{-}^4 g_{+}^4 \int\!\frac{d l_1}{(2\pi)^3}
 \int\!\frac{d l_2}{(2\pi)^3}
 D(l_1+\tfrac{q}{2})D(l_2-l_1)D(\tfrac{q}{2}-l_2) \notag \\
 & \times
 \Pi_{4,-}^{\mathrm{sym}}(\tfrac{q}{2}-l_2,l_2-l_1,l_1+\tfrac{q}{2},-q) \,
 \Pi_{4,+}(q,-\tfrac{q}{2}-l_1,l_1-l_2,l_2-\tfrac{q}{2}) .
\end{align}
The symmetrized 4-point loop $\Pi_{4,-}^{\mathrm{sym}}$ is defined by the sum of the six distinct permutations (without normalization factors such as $1/6$).
The individual Feynman diagrams corresponding to these permutations are depicted in Fig.~\ref{fig:fourloopdetail}.
Note that $g_{-}^4 g_{+}^4 = 1$.
We set the external momentum to $q=(0,0,q_y)$ and analyze the behavior for small $q_y$.
Since the external frequency is zero, the replacement of all loop frequencies by its negative produces the complex conjugate such that $\Pi^{(4e)}$ is real. In the same way we conclude that the integral is invariant under $q_y \rightarrow -q_y$, which means that the respective choice which N-point loop is on the plus and on the minus patch can be accounted for by a factor of 2.
\begin{figure}[tb]
\begin{center}
\includegraphics[width=10cm]{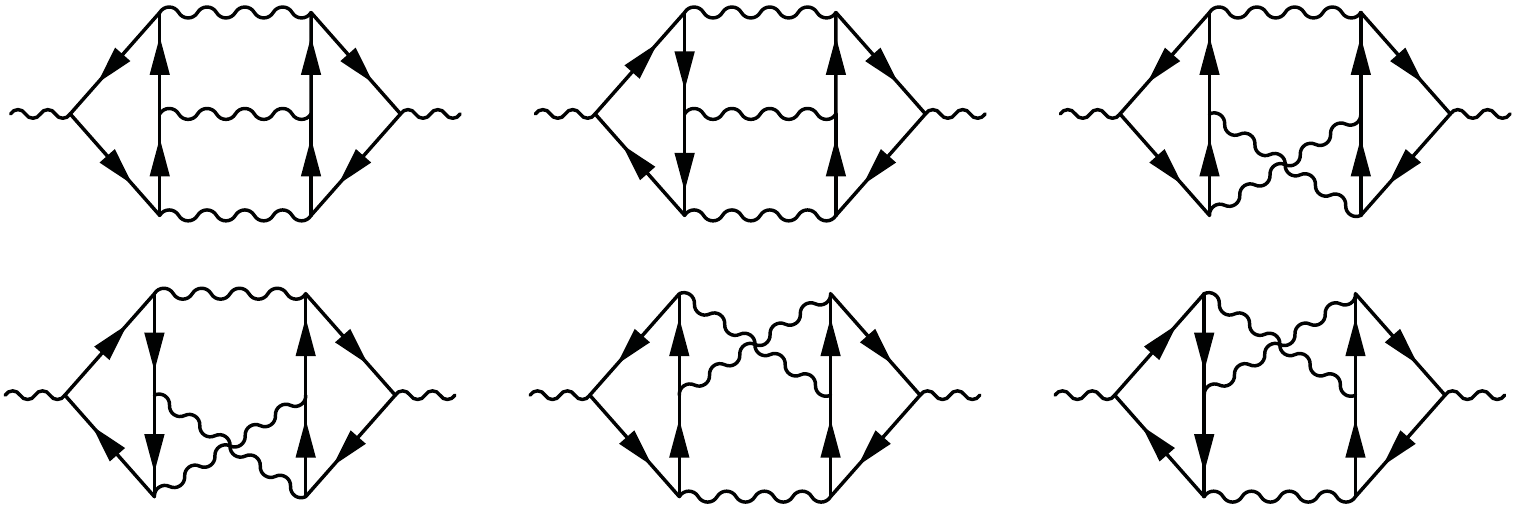}
\end{center}
\caption{The six four-loop contributions with two 4-point fermion loops contributing to the boson self-energy. All diagrams can be grouped into one by symmetrizing one of the 4-point loops.}
\label{fig:fourloopdetail}
\end{figure}

The 4-point loop is given by Eq.~(\ref{loopform}).
Only a few more simplifications are possible.
With some effort it is possible to perform the integrals in $l_{1x}$ and $l_{2x}$ by residues. However, this leads to a proliferation of terms and does not reduce the numerical effort required for the remaining integrations. Hence, we perform the integration over the bosonic variables $l_1$ and $l_2$ fully numerically.
The frequency integration within the fermion loops has to be done numerically, too.

We introduce stretched spherical integration variables by substituting 
$l_{i0} = r^3 \tilde l_{i0}$, $l_{ix} = r^2 \tilde l_{ix}$, and
$l_{iy} = r \tilde l_{iy}$, where $r$ runs from $0$ to $\Lambda$, and the tilde-variables are confined to a unit sphere.
Scaling out the $r$-dependence, the integral can then be written in the form
\begin{align}
 \Pi^{(4e)}(q_y) = \frac{q_y^2}{e^2} \int_0^{\Lambda} \frac{dr}{r} 
 \int d\Omega \, F(\Omega,q_y/r),
\end{align}
where $\Omega$ denotes the integration over the sphere.
A logarithmic divergence is then signaled by a nonzero result of the surface integral over $\Omega$ for $|q_y| \ll r$, and the prefactor of the divergence is given by
\begin{align}
 C^{(4e)} = \lim_{\tilde q_y \to 0} \int d\Omega \, F(\Omega,\tilde q_y) .
\end{align}

$\Pi_{4,s}$ itself contains 5 terms from the sum over $i<j$ (one term vanishes), the product $\Pi_{4,+}\Pi_{4,-}^{\mathrm{sym}}$ then contains $5\times30=150$ summands, each of which consists of a product of four $J_{ijk}$.
The integrand contains a large number of (integrable) poles.
We employed the computer algebra capabilities of \textsl{Wolfram Mathematica} and exported the results for an integration with the adaptive routine ``Divonne'' from \textsl{CUBA}~\cite{SHahn2005}.

\begin{figure}[htb]
\begin{center}
\includegraphics[width=8.5cm]{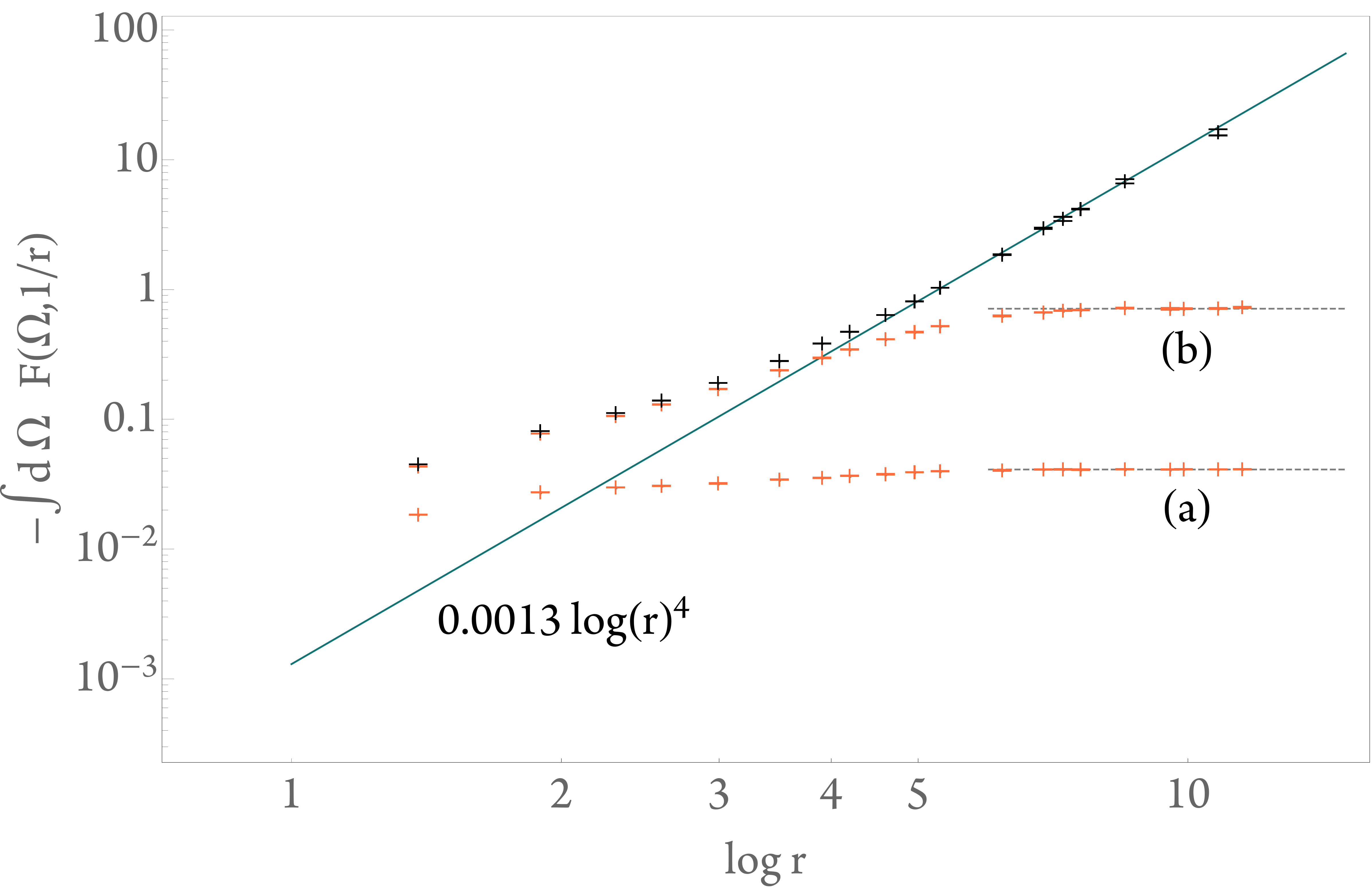}
\end{center}
\caption{Surface integral $\int d\Omega F(\Omega,q_y/r)$ for $q_y = 1$ as a function of $r$. The upper data points (black crosses) were computed with the $q_x$-independent boson propagator $D(q)$ as obtained from the RPA. The lower points (orange crosses) were obtained with an additional term proportional to $(q_x/q_y)^2$ with a prefactor $1/e^2$ (a) and $0.01/e^2$ (b) in the denominator of $D(q)$.}
\label{fig:divergence}
\end{figure}
The integral $\int d\Omega \, F(\Omega,\tilde q_y)$ turns out to be negative for small $|\tilde q_y|$ with increasing absolute value for decreasing $|\tilde q_y|$.
Instead of saturating, this value unexpectedly diverges as $[\log(\tilde q_y)]^4$ for $\tilde q_y \to 0$, see Fig.~S3. This implies that $\Pi^{(4e)}(q_y)$ diverges as $[\log(\Lam/|q_y|)]^5$ for $\Lam \to \infty$.
Adding a term proportional to $(q_x/q_y)^2$ to the denominator of $D(q)$, the divergence of $\int d\Omega \, F(\Omega,\tilde q_y)$ is regularized and one obtains a finite coefficient $C^{(4e)}$, as is also shown in Fig.~S3 for two distinct prefactors.

%%%%%%%%%%%%%%%%%%%%%%%%%%%%%%%%%%%%%%%%%%%%%%%%%%%%%%%%%%%%%%%%%%%%%%%%%%%%%%%%%%

%

\end{NoHyper}

\end{document}